\documentclass[preprint, showpacs,preprintnumbers,amsmath,amssymb,nofootinbib]{revtex4}
\usepackage{mathrsfs}
\usepackage{amsmath}
\usepackage{amsfonts}
\usepackage{latexsym}
\usepackage{amsfonts}
\usepackage{graphicx}
\usepackage{epsf}
\usepackage{dcolumn}
\usepackage{bm}

\newcommand{\bea}[1]{\begin{eqnarray}\label{#1}}
\newcommand{\eea}{\end{eqnarray}}

\def\gsim{ \lower .75ex \hbox{$\sim$} \llap{\raise .27ex \hbox{$>$}} }
\def\lsim{ \lower .75ex \hbox{$\sim$} \llap{\raise .27ex \hbox{$<$}} }

\begin{document}
 \title{Observational constraints on $f(T)$ theory}

\author{Puxun Wu $^{1,2}$
and  Hongwei Yu  $^{2,1}$ \footnote{Corresponding author: hwyu@hunnu.edu.cn}
}
\address
{$^1$ Center for Nonlinear Science and Department of Physics, Ningbo
University,  Ningbo, Zhejiang 315211, China\\
$^2$ Department of Physics and Key Laboratory of Low Dimensional
Quantum Structures and Quantum Control of Ministry of Education,
Hunan Normal University, Changsha, Hunan 410081, China
}

\begin{abstract}
The $f(T)$ theory, which is an extension of teleparallel, or torsion
scalar $T$, gravity, is recently proposed to explain the present
cosmic accelerating expansion with no need of dark energy. In this
Letter, we first perform the  statefinder  analysis and $Om(z)$
diagnostic to two concrete $f(T)$ models, i.e.,
 $f(T)=\alpha (-T)^n$ and  $f(T)=-\alpha
T(1-e^{p {T_0}/T})$,   and find that a crossing of phantom divide
line is impossible for both models. This is contrary to an existing
result where a crossing is claimed for the second model.  We, then,
study the constraints on them from the latest Union 2 Type Ia
Supernova (Sne Ia) set, the baryonic acoustic oscillation (BAO), and
the cosmic microwave background (CMB) radiation. Our results show
that at the $95\%$ confidence level
$\Omega_{m0}=0.272_{-0.032}^{+0.036}$, $n=0.04_{-0.33}^{+0.22}$  for
Model 1 and $\Omega_{m0}=0.272_{-0.034}^{+0.036}$,
$p=-0.02_{-0.20}^{+0.31}$ for Model 2.  A comparison of these two
models with the $\Lambda$CDM by the $\chi^2_{Min}/dof$ (dof: degree
of freedom) criterion indicates that $\Lambda$CDM is still favored
by observations.  We also study the evolution  of the equation of
state for the effective dark energy in the theory and find that Sne
Ia favors a phantom-like dark energy, while Sne Ia + BAO + CMB
prefers  a quintessence-like one.

\end{abstract}
 \pacs{04.50.Kd, 98.80.-k}
\maketitle

\section{Introduction}\label{sec1}
Various cosmological observations, including the Type Ia
Supernova~\cite{Riess1998}, the cosmic microwave background
radiation~\cite{Spergel2003} and the large scale
structure~\cite{Tegmark2004, Eisenstein2005}, et al., have revealed
that the universe is undergoing an accelerating expansion and it
entered this accelerating phase only in the near past. This
unexpected observed phenomenon poses  one of the most puzzling
problems in cosmology today. Usually, it is assumed that there
exists, in our universe, an exotic energy component with negative
pressure, named dark energy, which dominates the universe and drives
it to an accelerating expansion at recent times. Many candidates of
dark energy have been proposed, such as the cosmological constant,
quintessence, phantom, quintom as well as the (generalized)
Chaplygin gas, and so on. However, alternatively, one can take this
observed accelerating expansion as a signal of the breakdown of our
understanding to the laws of gravitation and, thus, a modification
of the gravity theory is needed. One of the most popular modified
gravity models is obtained by generalizing  the spacetime curvature
scalar $R$ in the Einstein-Hilbert action in  general relativity to
a general function of $R$. The theory so obtained  is called as  the
$f(R)$ theory (see \cite{Felice2010} for recent review).

Recently, a new modified gravity by extending the teleparallel
gravity~\cite{Einstein1930} is proposed to account for the present
accelerating expansion~\cite{Bengochea2009, Linder2010, Wu2010,
Bamba2010b}. Differing from general relativity using the Levi-Civita
connection, in teleparallel gravity,  the Weitzenb\"{o}ck connection
is used. As a result, the spacetime has only torsion and thus is
curvature-free. Similar to general relativity where the action is a
curvature scalar, the action of teleparallel gravity is a torsion
scalar $T$. In analogy to the $f(R)$ theory, Bengochea and Ferraro
suggested, in Ref.~\cite{Bengochea2009}, a new model, named $f(T)$
theory, by generalizing the action of teleparallel gravity, and
found that it can explain the observed acceleration of the universe.
Let us also note here that  models based on modified teleparallel
gravity may also provide an alternative to
inflation~\cite{Ferraro2007, Ferraro2008}.  Another advantage the
generalized $f(T)$ torsion theory has is that its field equations
are second order as opposed to the fourth order equations of f(R)
theory. More recently, Linder proposed two new $f(T)$ models to
explain the present cosmic accelerating expansion~\cite{Linder2010}
and found that the $f(T)$ theory can unify a number of interesting
extensions of gravity beyond general relativity.  In this Letter, we
plan to first perform a statefinder analysis and an $Om$ diagnostic
to these models  and then discuss the constraints on them from the
latest observational data, including the Type Ia Supernovae released
by the Supernova Cosmology Project Collaboration, the baryonic
acoustic oscillation from the spectroscopic Sloan Digital Sky
Survey, and the cosmic microwave background radiation from Wilkinson
Microwave Anisotropy Probe seven year observation. We find that for
both models the crossing of the $-1$ line is impossible. This  is
consistent with what obtained in Ref.~\cite{Bamba2010b}, but in
conflict with the result obtained in Ref.~\cite{Linder2010} where a
crossing is found for the exponential model.

\section{The $f(T)$ theory}
In this section, following Refs.~\cite{Bengochea2009, Linder2010},
we briefly review the $f(T)$ theory. We start with  teleparallel
gravity where the action is the torsion scalar $T$ defined as
 \begin{eqnarray}\label{ST}
 T\equiv S^{\;\;\mu\nu}_\sigma T^\sigma_{\;\;\mu\nu}\;,
 \end{eqnarray}
where $T^\sigma_{\;\;\mu\nu}$ is the torsion tensor
\begin{eqnarray}
T^\sigma_{\;\;\mu\nu}\equiv e^\sigma_A (\partial_\mu e^A_\nu- \partial_\nu e^A_\mu )\;,
\end{eqnarray}
and
\begin{eqnarray}
S^{\;\;\mu\nu}_\sigma\equiv\frac{1}{2}(K^{\mu\nu}_{\;\;\;\;\sigma}+\delta^\mu_\sigma T^{\alpha \nu}_{\;\;\;\;\alpha}-\delta^\nu_\sigma T^{\alpha \mu}_{\;\;\;\;\alpha})\;.
\end{eqnarray}
Here $e^A_\mu$ is the orthonormal tetrad component, where $A$ is an
index running over $0,1,2,3$ for the tangent space of the manifold,
while $\mu$, also running over   $0,1,2,3$, is the coordinate index
on the manifold.   The spacetime metric is related to $e^A_\mu$
through
\begin{eqnarray}
g_{\mu\nu}=\eta_{AB}e^A_\mu e^B_\nu\;,
\end{eqnarray}
and $K^{\mu\nu}_{\;\;\;\;\sigma}$ is the contorsion tensor given by
\begin{eqnarray}
 K^{\mu\nu}_{\;\;\;\;\sigma}=-\frac{1}{2}(T^{\mu\nu}_{\;\;\;\;\sigma}-T^{\nu\mu}_{\;\;\;\;\sigma}-T_{\sigma}^{\;\;\mu\nu})
 \end{eqnarray}

By assuming a flat homogeneous and isotropic
Friedmann-Robertson-Walker  universe which is described by the
metric
\begin{eqnarray}
ds^2=dt^2-a^2(t)\delta_{ij}dx^idx^j\;,
\end{eqnarray}
where $a$ is the scale factor, one has, from Eq.~(\ref{ST}),
\begin{eqnarray}
T=-6 H^2\;,
\end{eqnarray}
with $H=\dot{a}a^{-1}$ being the Hubble parameter.

In order to explain the late time cosmic accelerating expansion
without the  need of dark energy, Linder, following
Ref.~\cite{Bengochea2009}, generalized the Lagrangian density in
teleparallel gravity by promoting  $T$ to be $T+f(T)$.  The modified
Friedmann equation then becomes
\begin{eqnarray}\label{MF}
H^2=\frac{8\pi G}{3}\rho-\frac{f}{6}-2H^2f_{T}\;,
\end{eqnarray}
\begin{eqnarray} \label{MF2}
(H^2)'=\frac{16\pi GP+6H^2+f+12H^2f_{T}}{24H^2f_{T T}-2-2f_{T}}\;,
\end{eqnarray}
where a prime denotes a derivative with respect to $\ln a$, $\rho$
is energy  density  and $P$ is the pressure. Here we assume that the
energy component in the universe is only matter with  radiation
neglected, thus $P=0$.

From Eqs.~(\ref{MF}, \ref{MF2}), we can define an effective dark
energy, whose  energy density and the equation of state can be
expressed, respectively, as
\begin{eqnarray}
\rho_{eff}=\frac{1}{16\pi G}(-f + 2Tf_{T})
\end{eqnarray}
\begin{eqnarray}
w_{eff}=-\frac{f/T-f_{T}+2Tf_{TT}}{(1+f_{T}+2Tf_{TT})(f/T-2f_{T})}\;.
\end{eqnarray}

Some models are proposed
in Refs.~\cite{Bengochea2009, Linder2010} to explain  the present
cosmic accelerating expansion, which satisfy the usual condition
$f/T\rightarrow 0$ at the high redshift in order to be consistent
with the primordial nucleosynthesis and cosmic microwave background
constraints.  Here we consider two models proposed by
Linder~\cite{Linder2010}:

$\bullet$ Model 1
\begin{eqnarray}
f(T)=\alpha (-T)^n\;.
\end{eqnarray}
Here $\alpha$ and $n$ are two model parameters. Using the modified Friedmann equation, one can obtain
\begin{eqnarray}
\alpha=(6H_0^2)^{1-n}\frac{1-\Omega_{m0}}{2n-1}\;,
\end{eqnarray}
where $\Omega_{m0}=\frac{8\pi G \rho(0)}{3H_0^2}$ is the
dimensionless matter density today. Substituting above expression
into the modified Friedmann equation and defining $E^2=H^2/H^2_0$,
one has
\begin{eqnarray}\label{Mod1Ez}
E^2(z)=\Omega_{m0}(1+z)^3+(1-\Omega_{m0}) E^{2n}\;.
\end{eqnarray}
 Let us note that this model has the same background evolution equation as some
phenomenological models~\cite{Dvali2003, Chung2000} and  it reduces
to the $\Lambda$CDM model when $n=0$,  and to the DGP
model~\cite{Dvali2000} when $n=1/2$. When $n=1$, the Friedmann
equation (Eq.~(\ref{MF})) can be rewritten as $H^2=\frac{8\pi
G}{3(1-\alpha)}\rho$, which is the same as that of a standard cold
dark matter (SCDM) model if we rescale the Newton's constant as
$G\rightarrow G/(1-\alpha)$.    Therefore, in order to obtain an
accelerating expansion, it is required that $n<1$.

$\bullet$ Model 2
\begin{eqnarray}
f(T)=-\alpha T(1-e^{p{T_0}/T})\;,
\end{eqnarray}
which is similar to a $f(R)$ model where an exponential dependence
on the  curvature scalar is proposed~\cite{Linder2009, Bamba2010}.
Using the modified Friedmann equation again,  we have
\begin{eqnarray}
\alpha=\frac{1-\Omega_{m0}}{1-(1-2p)e^p}\;,
\end{eqnarray}
and
\begin{eqnarray}\label{Mod2Ez}
E^2(z)=\Omega_{m0}(1+z)^3+(1-\Omega_{m0})\frac{ E^{2}-
E^{2}e^{p/E^2}+2pe^{p/E^2}}{1-(1-2p)e^p}\;.
\end{eqnarray}
It is easy to see that $p=0$ corresponds to the case of $\Lambda$CDM.

\section{statefinder analysis and $Om$ diagnostic}
In order to discriminate different dark energy models from each
other, Sanhi et al. proposed a geometrical diagnostic method by
adding higher derivatives of the scale factor~\cite{Sanhi}. In this
method, two parameters $(r, s)$, named statefinder parameters, are
used, which  are defined,  respectively, as
\begin{eqnarray}
r\equiv \frac{\dddot{a}}{aH^3} \;,
\end{eqnarray}
\begin{eqnarray}
s\equiv \frac{r-1}{3(q-1/2)} \;,
\end{eqnarray}
where $q\equiv -\frac{1}{H^2}\frac{\ddot{a}}{a}$ is the decelerating
parameter. Apparently, $\Lambda$CDM model corresponds to a point
$(1,0)$ in $(r, s)$ phase space. The statefinder diagnostic can
discriminate different models. For example, it can distinguish
quintom from other dark energy models~\cite{WuYu}.

The $Om(z)$ is a new diagnostic of dark energy proposed by Sahni et
al.~\cite{Sahni2008}.  It is defined as
\begin{equation}
Om(z)\equiv\frac{E^2(z)-1}{(1+z)^3-1}.
\end{equation}
Apparently, this diagnostic only  depends on the first derivative of
the luminosity $D_L(z)$ (see Eq.~(\ref{dl})). Thus, its advantage, as
opposed to the equation of state of dark energy, is that it  is less
sensitive to the observational errors and the present matter energy
density $\Omega_{m0}$.  One can use this diagnostic to discriminate
different dark energy  models by examining the slope of $Om(z)$ even
if the value of $\Omega_{m0}$ is not exactly known, since the
positive, null, or negative slopes correspond to $w<-1, w=-1$ or
$w>-1$, respectively.

\begin{figure}[htbp]
\includegraphics[width=5cm]{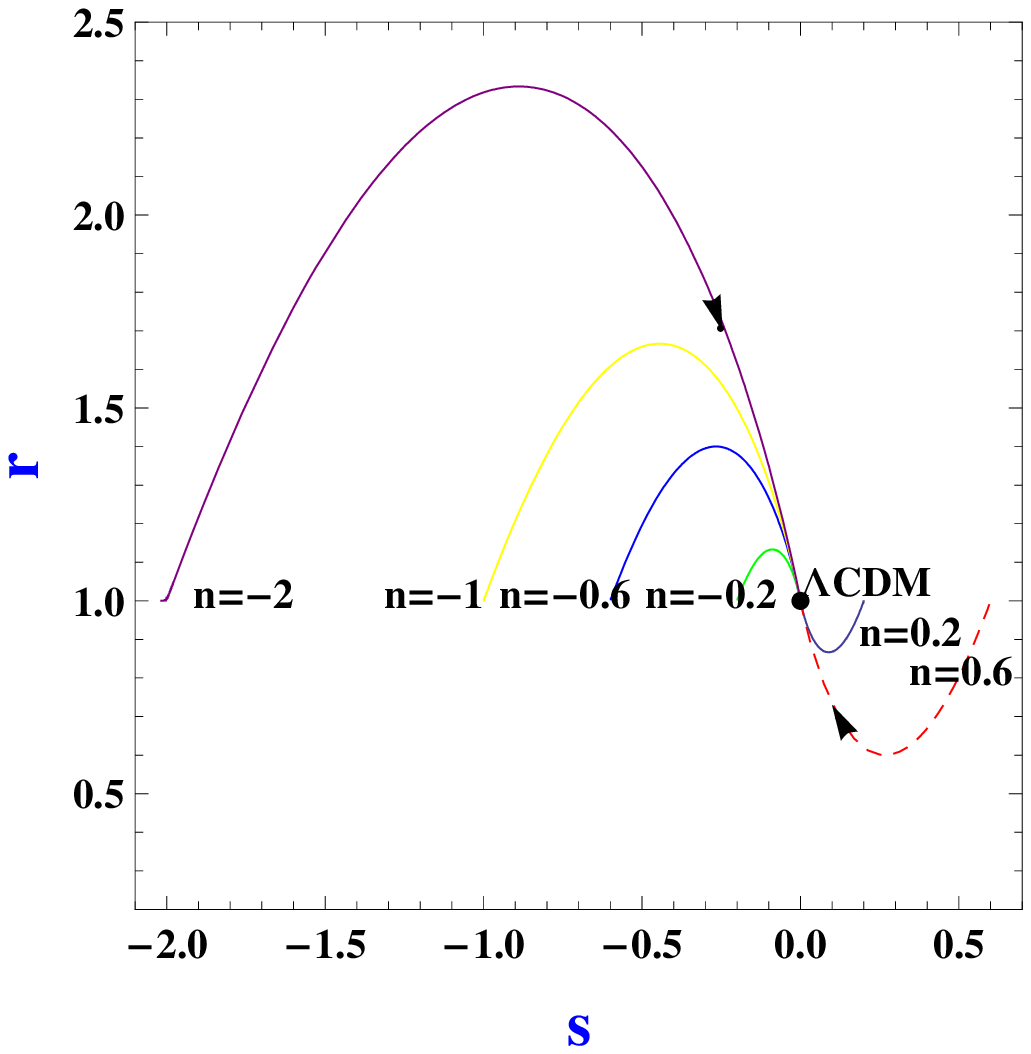}\includegraphics[width=5cm]{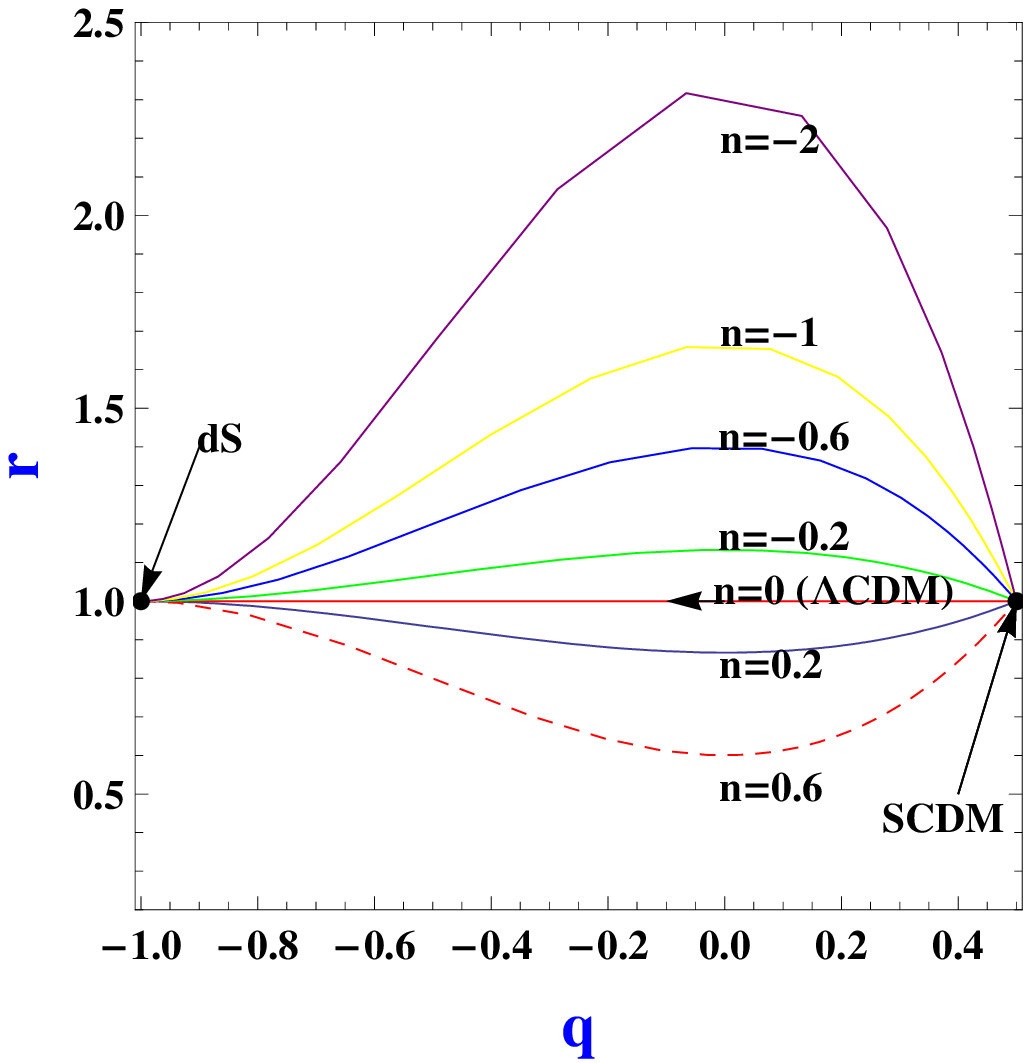}\includegraphics[width=5cm]{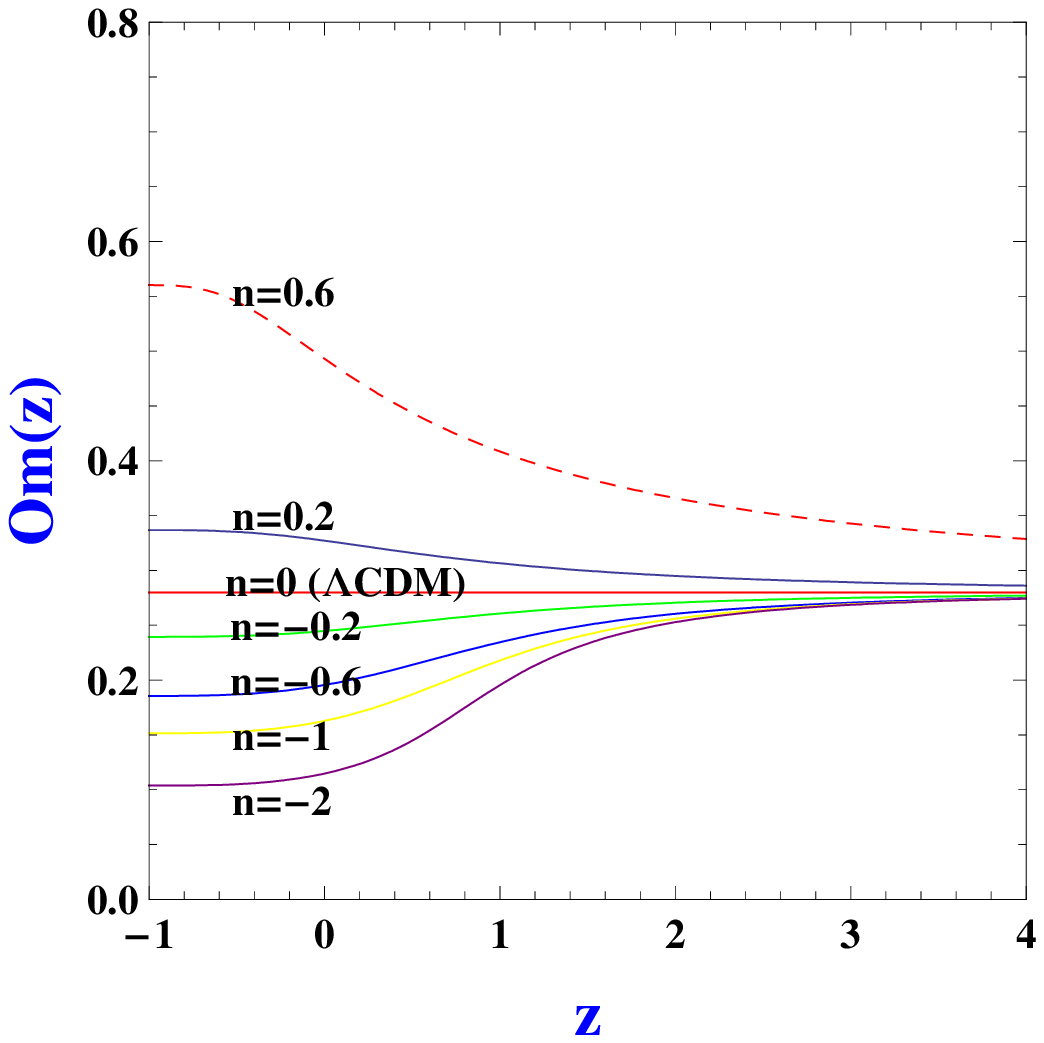}
\caption{\label{Figs1} The evolutionary curves of statefinder pair
$(r, s)$ (left), pair $(r, q)$ (middle) and $Om(z)$ (right) for
Model 1 with $\Omega_{m0}=0.278$. }
\end{figure}

\begin{figure}[htbp]
\includegraphics[width=5cm]{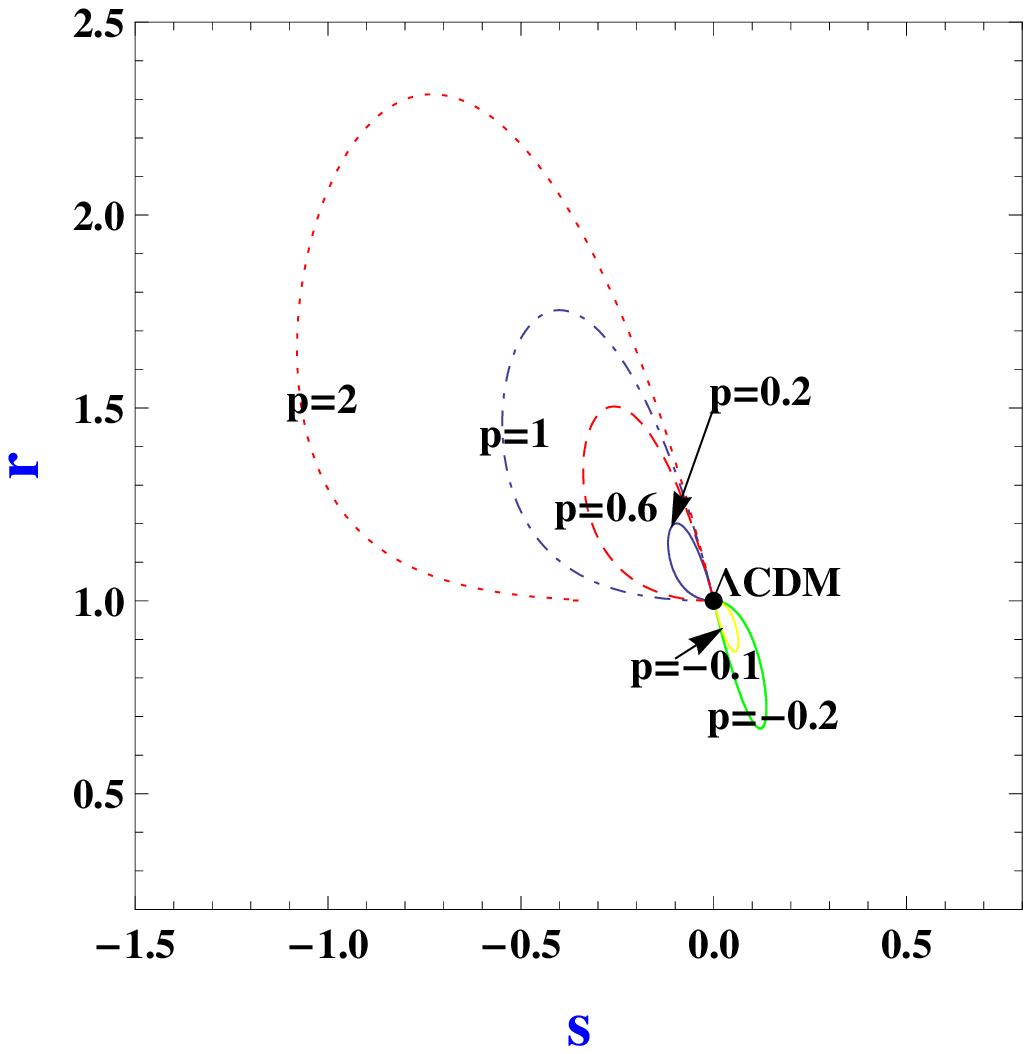}\includegraphics[width=5cm]{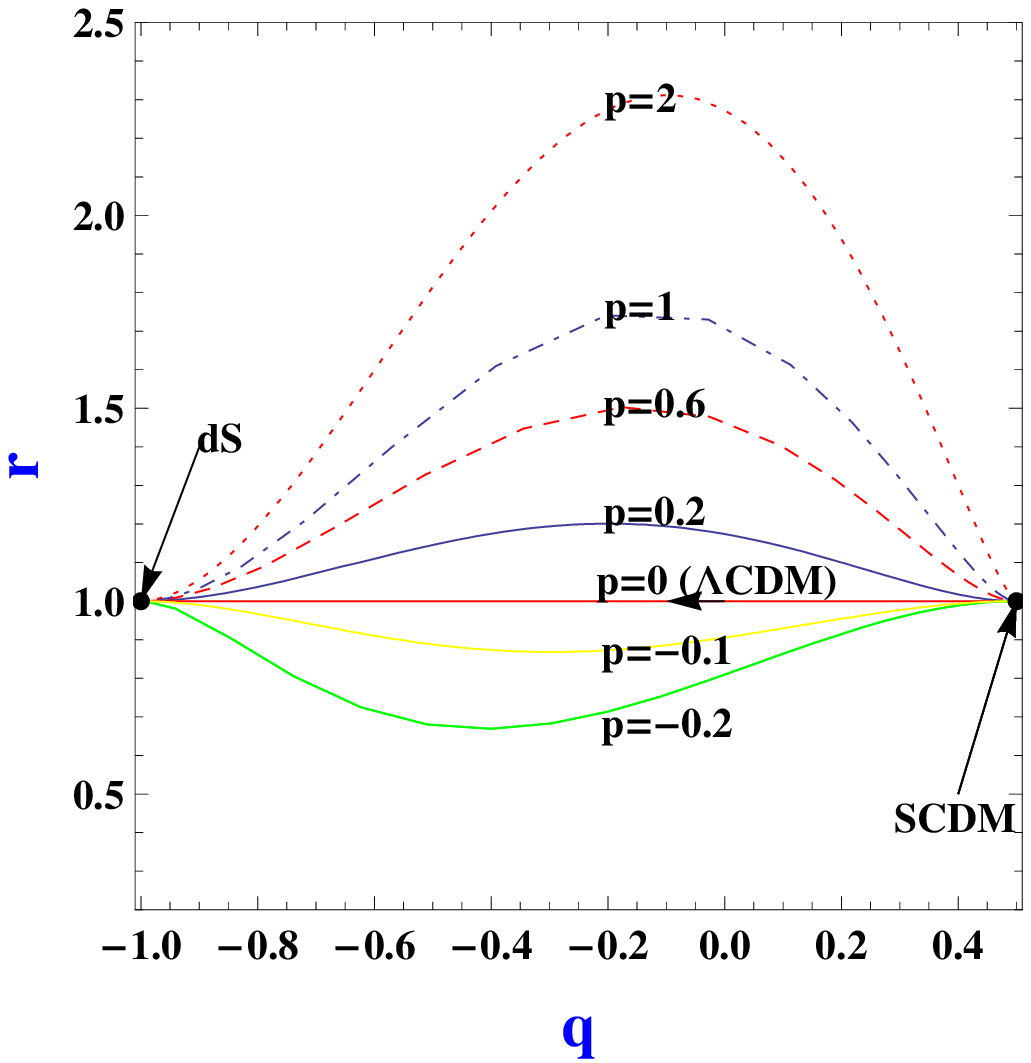}\includegraphics[width=5cm]{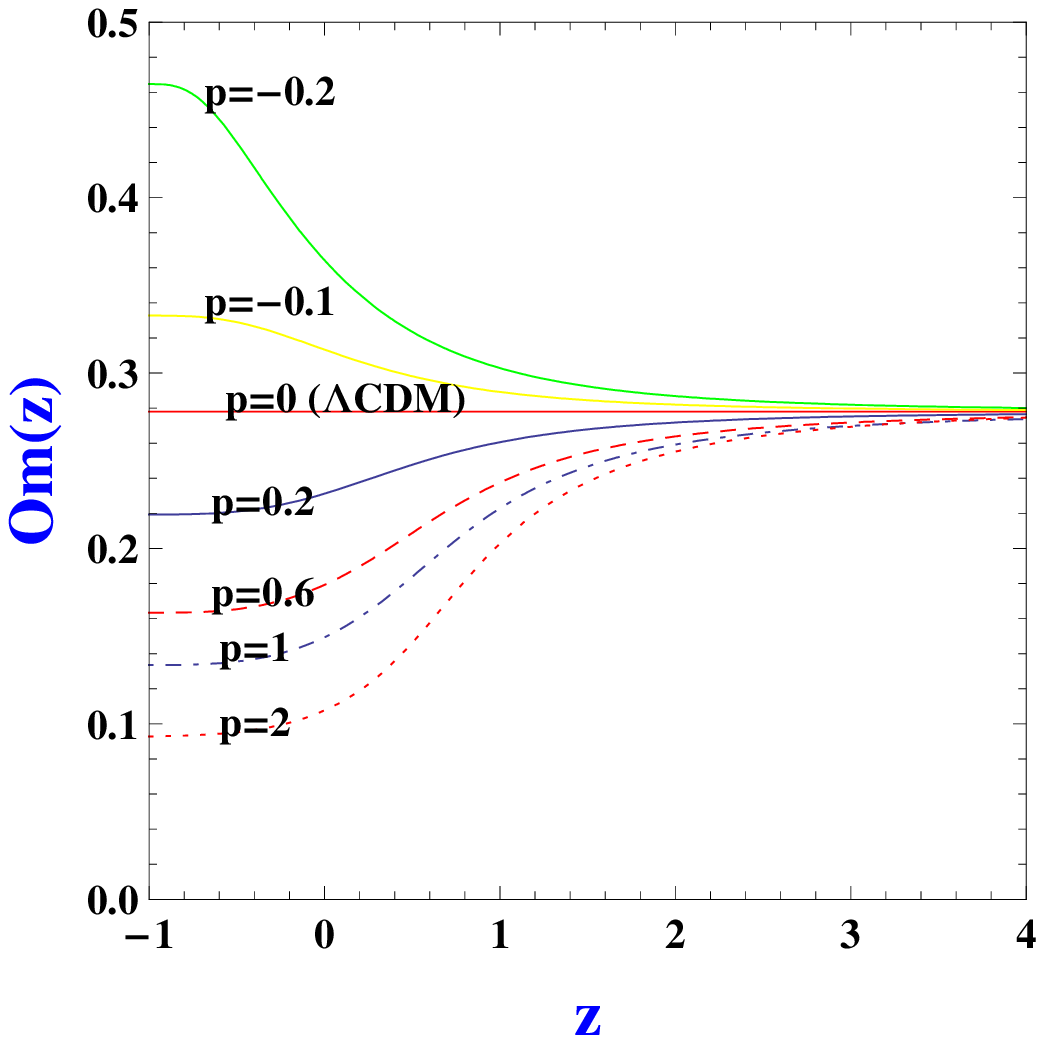}
\caption{\label{Figs2} The evolutionary curves of statefinder pair
$(r, s)$ (left), pair $(r, q)$ (middle) and $Om(z)$ (right) for
Model 2 with $\Omega_{m0}=0.278$. }
\end{figure}

Here, we perform the statefinder and $Om$ diagnostics to two $f(T)$
models, i.e.,  Model 1 and Model 2 given in the previous section. In
Figs.~(\ref{Figs1}) and (\ref{Figs2}), we show the diagnostic
results with $\Omega_{m0}=0.278$, which is the best fit value
obtained from Sne Ia and BAO with a model independent
method~\cite{Wu2008} and is also consistent with the result in the
next section of the present Letter. The left panels show the
evolutionary curves of statefinder pair $(r,s)$,  the middle panels
are the evolutionary curves of pair $(r, q)$, and the right panels
are the $Om(z)$ diagnostic. Although,  both Model 1 and Model 2
evolve from the SCDM to a de Sitter (dS) phase as one can see from
the middle panels of these figures, the effective dark energy  for
Model 2 with $p\neq 0$
 is similar to a cosmological
constant both in the high redshift regimes and in the future, while
for Model 1 with $n\neq 0$ this similarity occurs only in the
future.

As demonstrated in Ref.~\cite{Ali2010}, for a simple power law
evolution of the scale factor $a(t)\simeq t^{2/3 \gamma}$, one has
$r=(1-3\gamma)(1-3\gamma/2)$ and $s=\gamma$. Thus, a phantom-like
dark energy corresponds to $s<0$, a quintessence-like dark energy to
$s>0$, and an evolution from phantom to quintessence or inverse is
given by a crossing of the point $(1, 0)$ in $(r, s)$ phase plane. A
crossing of phantom divide line is also represented by a crossing of
the red solid line ($\Lambda$CDM) in middle panels ($(r, q)$ plane)
of Figs.~(\ref{Figs1}, \ref{Figs2}). Therefore, we find, from the
left and middle panels of Figs.~(\ref{Figs1}, \ref{Figs2}), that
$n>0$ (Model 1) or $p<0$ (Model 2) $f(T)$ corresponds to a
quintessence-like dark energy model, while $n<0$ (Model 1) or $p>0$
(Model 2) corresponds to a phantom-like one. A crossing of the
phantom divide line  is impossible for Model 1 and Model 2. These
results are also confirmed by the $Om(z)$ analysis given in the
right panels. In order to further confirm our results, we redo our
analysis with other values of $\Omega_{m0}$, such as
$\Omega_{m0}=0.2$ or $0.5$, and find  that the  result remains
unchanged. Thus, we conclude that the phantom divide line is not
crossed for both models. This is in conflict with what given in
Ref.~\cite{Linder2010} where a crossing of the phantom line is found
for  Model 2.

\section{observational constraints}
The constraints on model parameters of Model 1 and Model 2 will be
discussed,   respectively, in this section. Three different kinds of
observational data, i.e.,  the Type Ia supernovae (Sne Ia), the
baryonic acoustic oscillation (BAO) from the spectroscopic Sloan
Digital Sky Survey (SDSS) and the cosmic microwave background (CMB)
radiation from Wilkinson Microwave Anisotropy Probe (WMAP), will be
used in order to break the degeneracy between the model parameters.
The fitting methods are summarized in the Appendix.

For the  Sne Ia data, we use the Union 2 compilation released by the
Supernova  Cosmology Project collaboration
recently~\cite{Amanullah2010}.  Calculating the ${\chi}^2_{Sne}$, we
find that, for Model 1, the best fit values occur at
$\Omega_{m0}=0.302$, $n=-0.18$ with $\chi^2_{Min}=543.953$, whereas,
for Model 2, $\Omega_{m0}=0.279$, $p=0.08$ with
$\chi^2_{Min}=543.369$.

\begin{figure}[htbp]
\includegraphics[width=6cm]{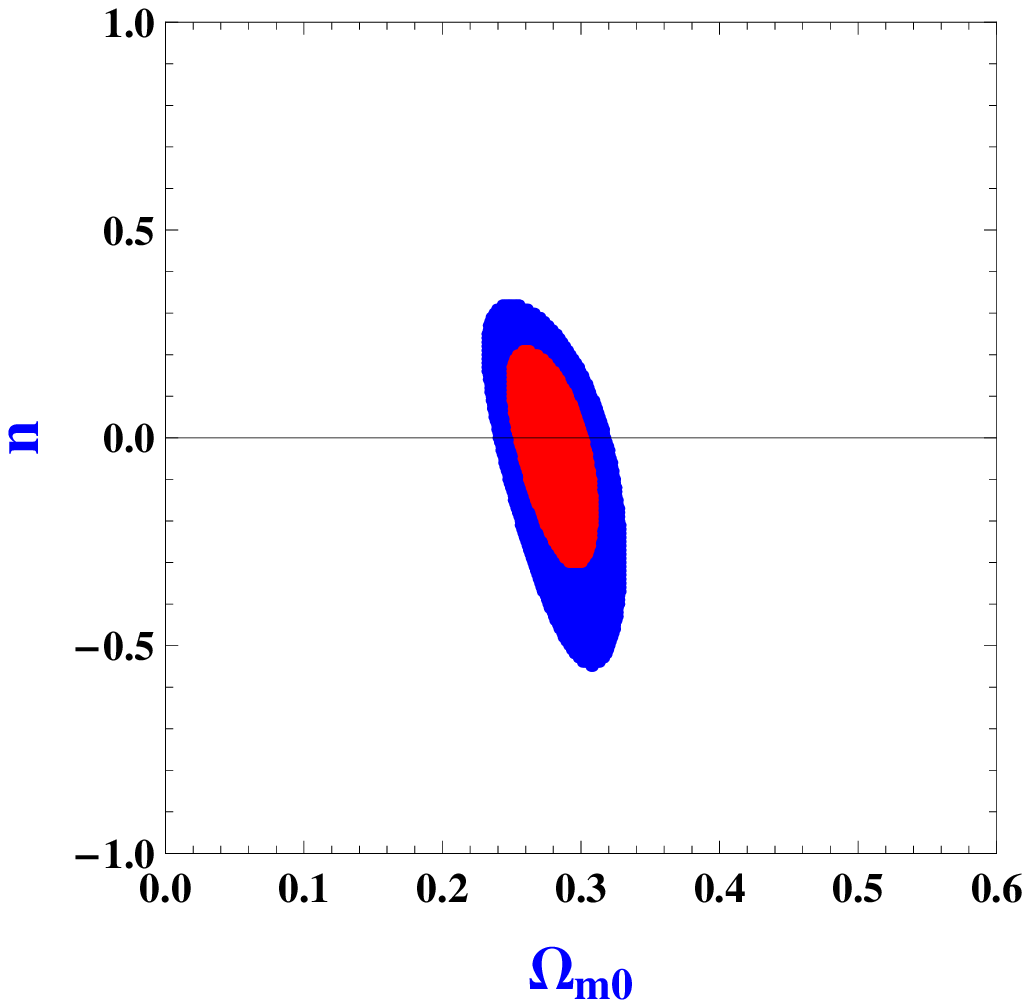}\quad\includegraphics[width=6cm]{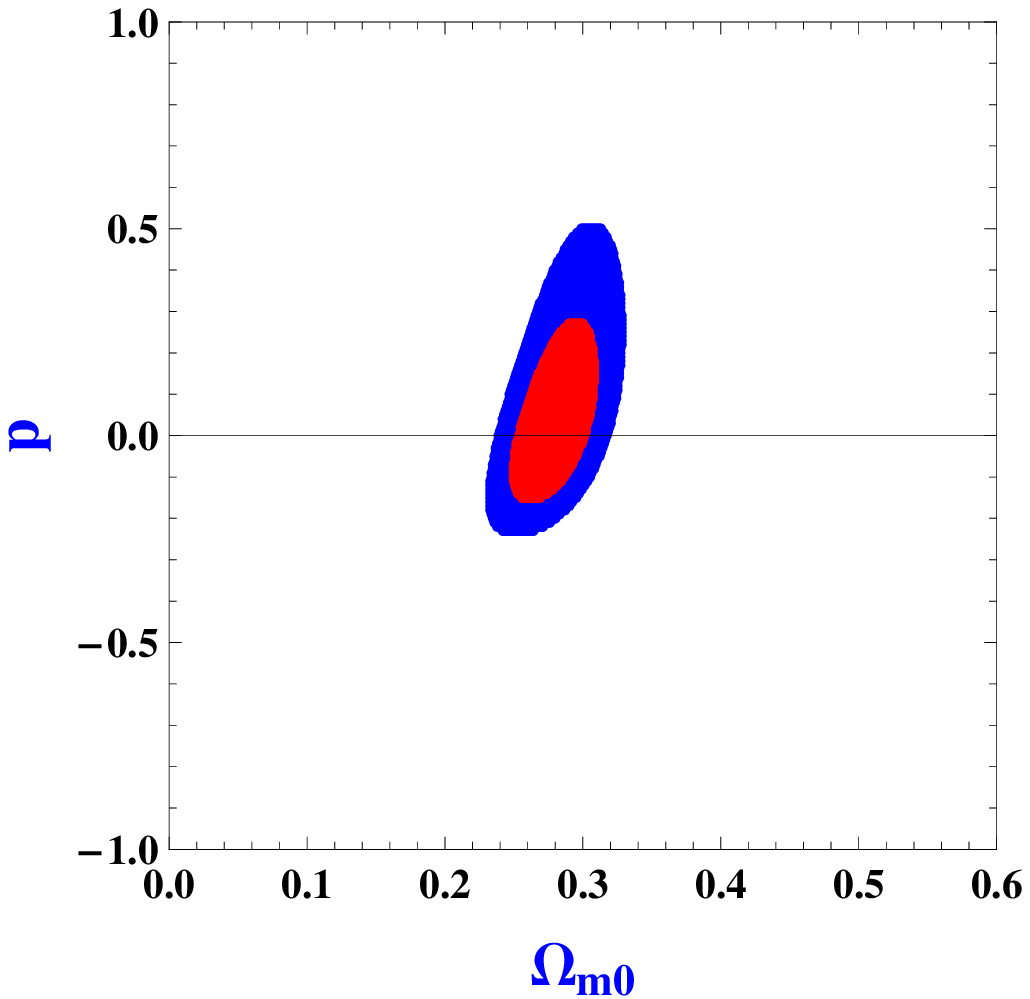}
 \caption{\label{Fig1} The constraints on  Model 1 (left) and Model 2 (right) from  Sne Ia + BAO.
The red and blue+red regions correspond to $1-\sigma$ and $2-\sigma$ confidence regions, respectively.}
\end{figure}

Then, we consider the constraints from the  BAO data. The parameter
$A$ given by the BAO peak in the distribution of SDSS luminous red
galaxies~\cite{Eisenstein2005} is used. The constraints from Sne
Ia+BAO are given by minimizing $\chi^2_{Sne}+\chi^2_{BAO}$. The
results are $\Omega_{m0}=0.279_{-0.047}^{+0.050}$,
$n=-0.01_{-0.54}^{+0.31}$ (at the $95\%$ confidence level) with
$\chi^2_{Min}=542.978$ for Model 1 and
$\Omega_{m0}=0.278_{-0.045}^{+0.050}$, $p=0.02_{-0.24}^{+0.48}$ (at
the $95\%$ confidence level) with $\chi^2_{Min}=543.383$ for Model
2. The contour diagrams are shown in Fig.~(\ref{Fig1}).

\begin{figure}[htbp]
\includegraphics[width=6cm]{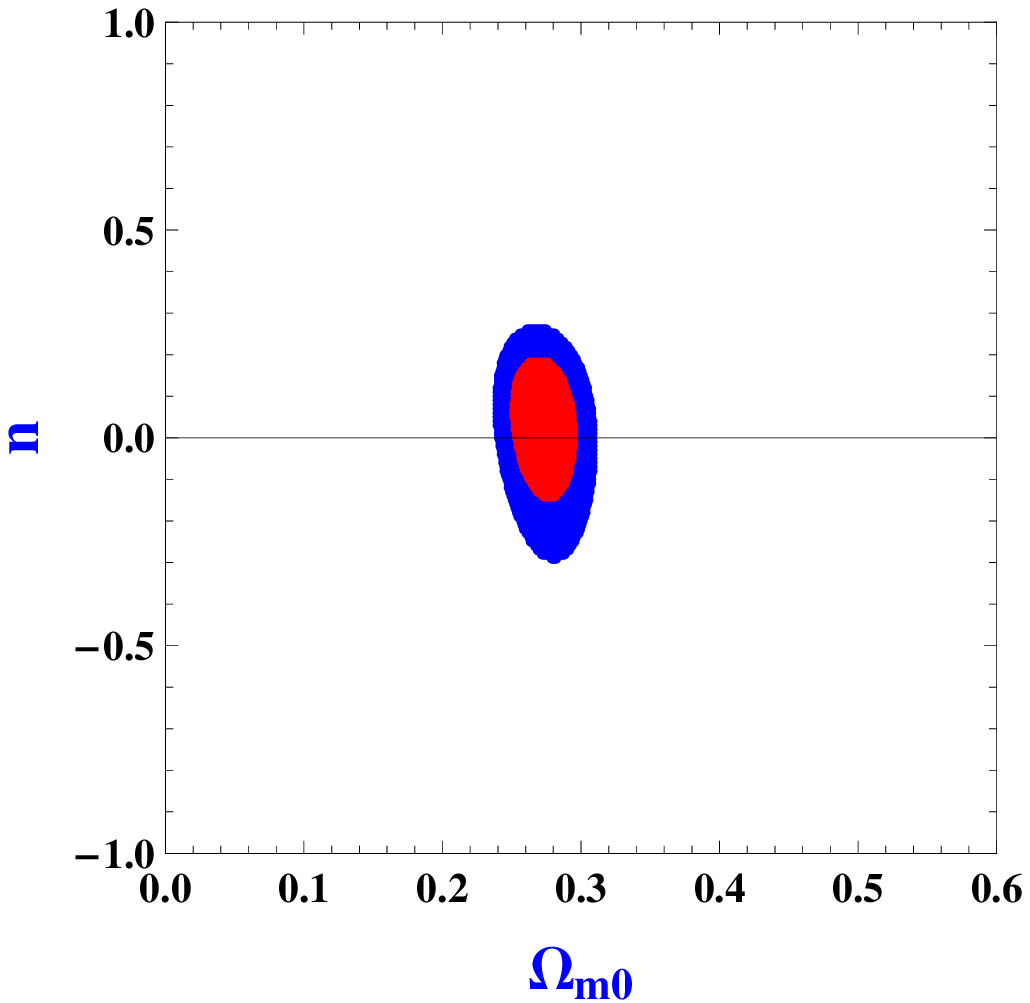}\quad\includegraphics[width=6cm]{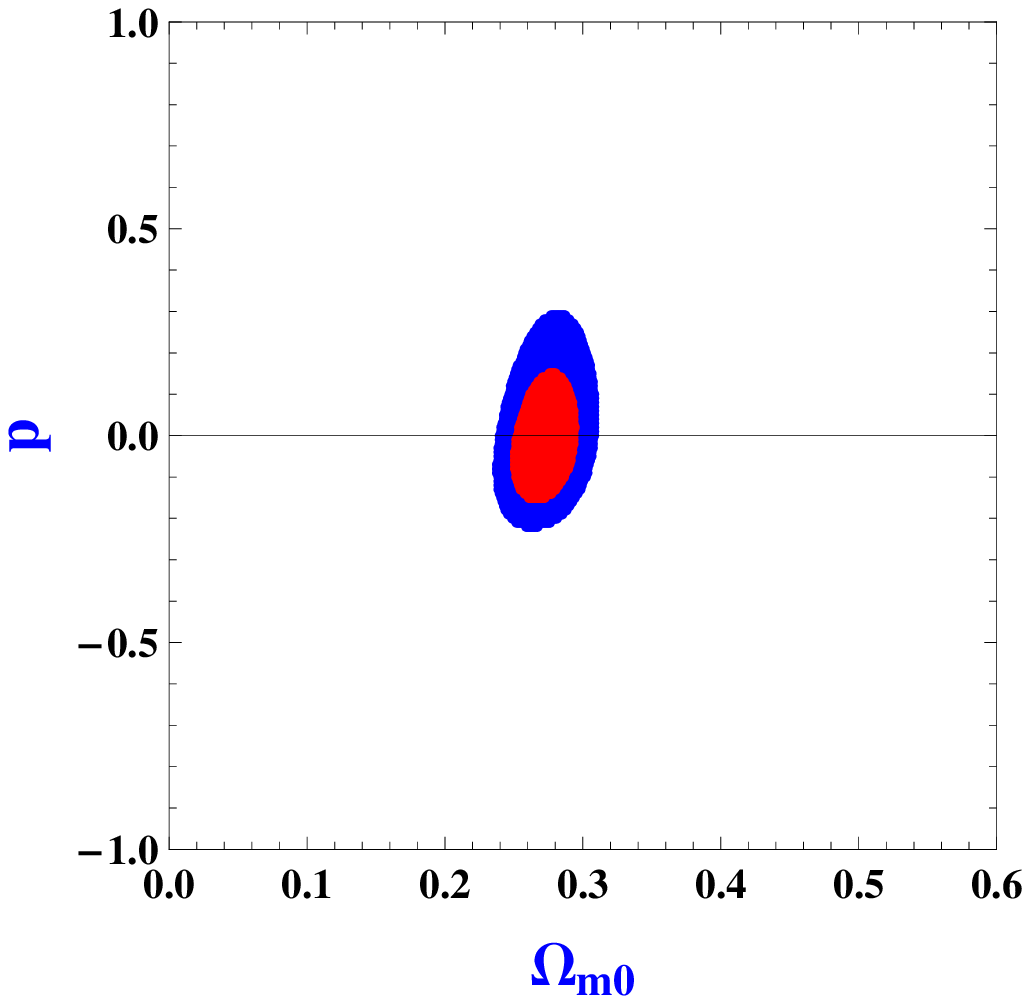}
 \caption{\label{Fig2} The constraints on  Model 1 (left) and Model 2 (right) from  Sne Ia + BAO + CMB.
The red and blue+red regions correspond to $1-\sigma$ and $2-\sigma$ confidence regions, respectively.}
\end{figure}

Furthermore, the CMB data is added in our analysis. The CMB shift
parameter $R$~\cite{Wang2006, Bond1997} is used. The constraints
from Sne Ia + BAO + CMB  are given by
$\chi^2_{all}=\chi^2_{Sne}+\chi^2_{BAO}+\chi^2_{CMB}$.
Fig.~(\ref{Fig2}) shows the results. We find that, at the $95\%$
confidence level, $\Omega_{m0}=0.272_{-0.032}^{+0.036}$,
$n=0.04_{-0.33}^{+0.22}$  with $\chi^2_{Min}=543.168$ for Model 1
and
  $\Omega_{m0}=0.272_{-0.034}^{+0.036}$,
$p=-0.02_{-0.20}^{+0.31}$ with $\chi^2_{Min}=543.631$ for Model 2.

With the observational data considered above, we also discuss the
constraints on the $\Lambda$CDM and the results are
$\Omega_{m0}=0.277_{-0.038}^{+0.040}$ with $\chi^2_{Min}=543.400$
(Sne Ia + BAO)   and $\Omega_{m0}= 0.276_{-0.036}^{+0.032}$ with
$\chi^2_{Min}=543.745$ (Sne Ia + BAO + CMB) at the $95\%$ confidence
level. A summary of constraint results on Model 1, Model 2 and
$\Lambda$CDM is given in Table (1). From Figs.~(\ref{Fig1},
\ref{Fig2}) and Table (1), one can see that the $\Lambda$CDM
(corresponding to $n=0$ for Model 1 and $p=0$ for Model 2) is
consistence with the observations at the $68\%$ confidence level,
while the DGP model (corresponds to $n=1/2$ for Model 1) is ruled
out at the $95\%$ confidence level. Meanwhile, using the
$\chi^2_{Min}/dof$ (dof: degree of freedom) criterion, we find that
the $\Lambda$CDM is favored by observations.

\begin{table}[!h]
\tabcolsep 0.5pt \caption{\label{Tab1} Summary of the constraint on
model  parameters and $\chi^2_{Min}/dof$. In the table S+B+C
represents Sne Ia + BAO + CMB. }\vspace*{-12pt}
\begin{center}
\begin{tabular}{|c|c|c|c|c|c|c|c|c|c|}
  \hline
      & \multicolumn{3}{|c|}{Model 1}         & \multicolumn{3}{|c|}{Model 2}  & \multicolumn{2}{|c|}{$\Lambda$CDM}\\ \cline{2-9}
           & $\Omega_{m0}$ & $n$   &$\chi^2_{Min}/dof$    & $\Omega_{m0}$  & $p$ &$\chi^2_{Min}/dof$ & $\Omega_{m0}$ &$\chi^2_{Min}/dof$  \\ \hline
 Sne+BAO  & $0.279_{-0.047}^{+0.050}$ &  $-0.01_{-0.54}^{+0.31} $&  0.974    & $0.278_{-0.045}^{+0.050}$  & $0.02_{-0.24}^{+0.48}$ & 0.975 & $0.277_{-0.038}^{+0.040}$ & 0.973\\ \hline
 S+B+C  & $0.272_{-0.032}^{+0.036}$  & $0.04_{-0.33}^{+0.22} $ & 0.975  & $0.272_{-0.034}^{+0.036}$ & $-0.02_{-0.20}^{+0.31}$  & 0.976 &$ 0.276_{-0.036}^{+0.032}$ &0.974\\
  \hline
\end{tabular}
       \end{center}
       \end{table}

\begin{figure}[htbp]
\includegraphics[width=6cm]{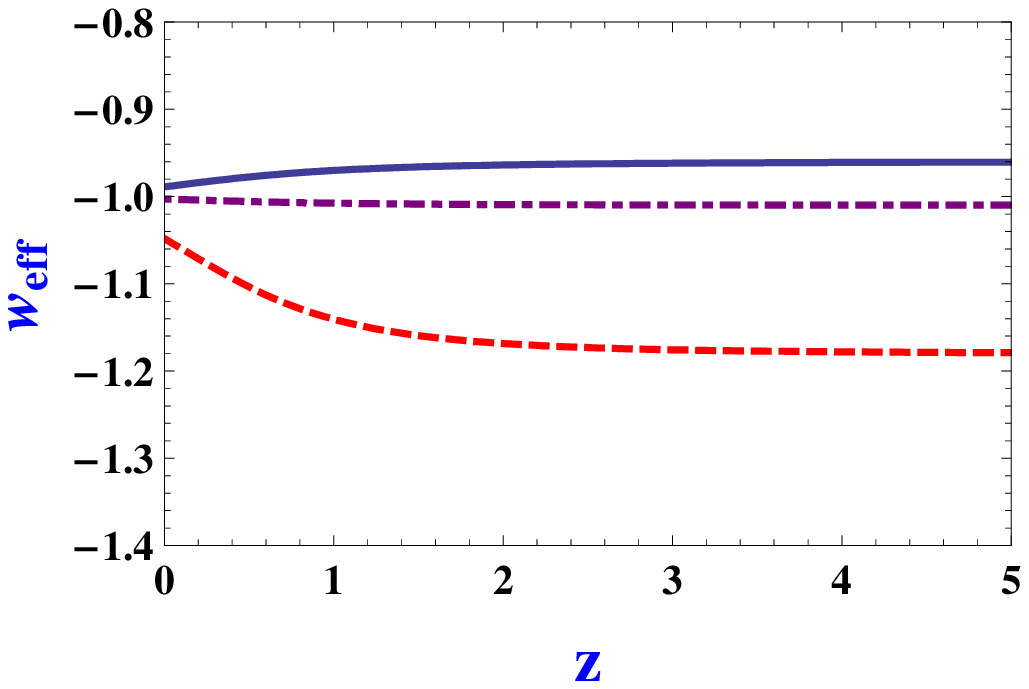}\quad\includegraphics[width=6cm]{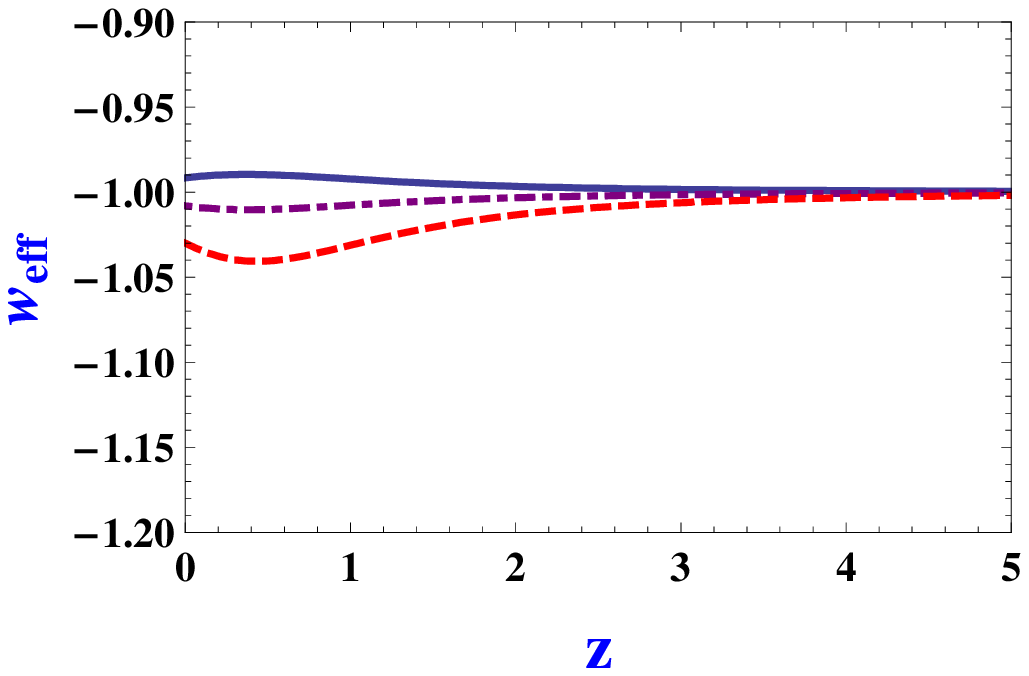}
 \caption{\label{Fig3} The evolutionary curves of the equation of state for the effective dark energy from  Model1 (left) and Model2 (right). The  model parameters are set at the best fit values. The dashed, dotdashed and solid  lines correspond to the constraints from Sne Ia,  Sne Ia + BAO, and Sne Ia + BAO + CMB, respectively.}
\end{figure}

In addition, we study the evolution of the equation of state for the
effective dark energy.  The results are shown in Fig.~(\ref{Fig3}).
The dashed, dotdashed  and solid lines show the evolutionary curves with the
model parameters at the best fit values from  Sne Ia, Sne Ia+BAO,
and Sne Ia+BAO+CMB, respectively. Apparently, Sne Ia favors a
phantom-like dark energy, while Sne Ia + BAO + CMB favor a
quintessence-like one.

\section{Conclusion}
Recently, the $f(T)$ gravity theory is proposed to explain the
present cosmic accelerating expansion without the need of dark
energy. In this Letter, we discuss firstly the statefinder
geometrical analysis and $Om(z)$ diagnostic to the $f(T)$ gravity.
Two concrete $f(T)$ models proposed by Linder~\cite{Linder2010} are
studied. From the $Om(z)$ diagnostic and the phase space analysis of
the statefinder parameters $(r,s)$ and pair $(s,p)$,  we find  that,
for both  Model 1 and Model 2, a crossing of the phantom divide line
is impossible, which conflicts with the result obtained in
Ref.~\cite{Linder2010} where a crossing is found for Model 2. We
then consider the constraints on Model 1 and Model 2 from the latest
Union 2 Type Ia Supernova set released by the Supernova Cosmology
Project collaboration, the baryonic acoustic oscillation observation
from the spectroscopic Sloan Digital Sky Survey Data Release galaxy
sample, and the cosmic microwave background radiation observation
from the seven-year Wilkinson Microwave Anisotropy Probe result. We
find that at the $95\%$ confidence level, for Model 1,
$\Omega_{m0}=0.272_{-0.032}^{+0.036}$, $n=0.04_{-0.33}^{+0.22}$ with
$\chi^2_{Min}=543.168$ and for Model 2,
$\Omega_{m0}=0.272_{-0.034}^{+0.036}$, $p=-0.02_{-0.20}^{+0.31}$
with $\chi^2_{Min}=543.631$. We also find that the $\Lambda$CDM
(corresponds to $n=0$ for Model 1 and $p=0$ for Model 2) is
consistence with observations at $1-\sigma$ confidence level and it
is favored by  observation through the $\chi^2_{Min}/dof$ (dof:
degree of freedom) criterion. However, the DGP model, which
corresponds to $n=1/2$ for Model 1,  is ruled out by observations at
the $95\%$ confidence level. Finally, we study the evolution of the
equation of state for the effective dark energy in the $f(T)$
theory. Our results show that Sne Ia favors a phantom-like dark
energy, while Sne Ia + BAO + CMB prefers a quintessence-like one.
The analysis of the current paper also indicates that the $f(T)$
theory can give the same background evolution as other models such
as $\Lambda$CDM,  although they have completely different
theoretical basis. Thus, it remains interesting to study other
aspects of $f(T)$ theory, such as the matter density growth,  which
may help us distinguish it from other gravity theories.

\begin{acknowledgments}

This work was supported in part by the National Natural Science
Foundation of China under Grants Nos. 10775050, 10705055, 10935013
and 11075083,  Zhejiang Provincial Natural Science Foundation of
China under Grant No. Z6100077, the SRFDP under Grant No.
20070542002, the FANEDD under Grant No. 200922, the National Basic
Research Program of China under Grant No. 2010CB832803, the NCET
under Grant No. 09-0144, the PCSIRT under Grant No. IRT0964,  and
K.C. Wong Magna Fund in Ningbo University.

\end{acknowledgments}

\appendix

\section{Data and Fitting Method}\label{app:obervations}

\subsection{Type Ia Supernovae }

Recently,  the
Supernova  Cosmology Project collaboration
~\cite{Amanullah2010} released the Union2 compilation, which consists of 557 Sne
Ia data points and is the largest published and spectroscopically
confirmed Sna Ia sample today. We use it to constrain the theoretical models in this paper. The results  can
be obtained by minimizing the $\hat{\chi}^2$ value of the distance
moduli
\begin{eqnarray}
\hat{\chi}^2_{Sne}=\sum_{i=1}^{557}\frac{[\mu_{obs}(z_i)-\mu_{th}(z_i)]^2}{\sigma_{u,i}^2}\;,
\end{eqnarray}
where  $\sigma_{\mu,i}^2$ are the errors due to the flux
uncertainties, intrinsic dispersion of Sne Ia absolute magnitude and
peculiar velocity dispersion. $\mu_{obs}$ is the observed distance
moduli and $\mu_{th}$ is the theoretical one, which is defined as
\begin{eqnarray}
\mu_{th}=5 \log_{10}D_L-\mu_0\;.
\end{eqnarray}
Here $\mu_0=5\log_{10}h+42.38$, $h=H_0/100 km/s/Mpc$, and the luminosity distance $D_L$ can be calculated by
\begin{eqnarray}\label{dl}
D_L\equiv(1+z)\int_0^z\frac{dz'}{E(z')}\;,
\end{eqnarray}
with $E(z)$ given in Eqs.~(\ref{Mod1Ez}, \ref{Mod2Ez}).  Since  $\mu_0$
(or $h$) is a nuisance parameter, we marginalize  over it by an
effective approach given in Ref.~\cite{Nesseris2005}. Expanding
$\hat{\chi}^2_{Sne}$ to $\hat{\chi}^2_{Sne}(\mu_0)=A
\mu_0^2-2B\mu_0+C$ with $A=\sum1/\sigma_{u,i}^2$,
$B=\sum[\mu_{obs}(z_i)-5 \log_{10}D_L]/\sigma_{u,i}^2$ and
$C=\sum[\mu_{obs}(z_i)-5 \log_{10}D_L]^2/\sigma_{u,i}^2$, one can
find that $\hat{\chi}^2_{Sne}$ has a minimum value at $\mu_0=B/A$,
which is given by
\begin{eqnarray} \chi^2_{Sne}=C-\frac{B^2}{A}\;.\end{eqnarray}
Thus, we can minimize ${\chi}^2_{Sne}$ instead of
$\hat{\chi}^2_{Sne}$ to obtain  constraints from Sne Ia.

\subsection{Baryon Acoustic Oscillation}
For BAO data,  the parameter $A$ given by the BAO peak in the distribution
of SDSS luminous red galaxies~\cite{Eisenstein2005} is used. The
results can be obtained by calculating:
\begin{eqnarray}
\chi^2_{BAO}=\frac{[A-A_{obs}]^2}{\sigma_A^2}
\end{eqnarray}
where $A_{obs}=0.469(n_s/0.98)^{-0.35}\pm 0.017$ with the scalar
spectral  index $n_s=0.963$ from the WMAP 7-year
data~\cite{Komatsu2010} and the theoretical value $A$ is defined as
\begin{eqnarray}
A\equiv \Omega_{m0}^{1/2}E(z_b)^{-1/3}\bigg[\frac{1}{z_b}\int_0^{z_b}\frac{dz'}{E(z')}\bigg]^{2/3}
\end{eqnarray}
with $z_b=0.35$.
\subsection{Cosmic Microwave Background}
 Since the CMB
shift parameter $R$~\cite{Wang2006, Bond1997}  contains the
main information of the observations of the CMB, it is used in our analysis. The WMAP7
data gives the observed value of $R$ to be $R_{obs}=1.725\pm
0.018$~\cite{Komatsu2010}. The corresponding theoretical value is
defined as
\begin{eqnarray}
R\equiv \Omega_{m0}^{1/2}\int_0^{z_{CMB}}\frac{dz'}{E(z')}\;,
\end{eqnarray}
where $z_{CMB}=1091.3$. Therefore, the constraints on model
parameters can be  obtained by fitting the observed value with   the
corresponding theoretical one of parameter $R$ through the following
expression
\begin{eqnarray}
\chi^2_{CMB}=\frac{[R-R_{obs}]^2}{\sigma_R^2}.
\end{eqnarray}

\end{document}